\documentstyle[prb,aps,multicol]{revtex}
  \renewcommand{\narrowtext}{\begin{multicols}{2} \global\columnwidth20.5pc}
  \renewcommand{\widetext}{\end{multicols} \global\columnwidth42.5pc}
  \newcommand{\wide}{\widetext \noindent \line(200,0){245} \line(0,1){3}\\}
  \newcommand{\narrow}{\begin{flushright}\mbox{\line(0,-1){3}$\! \!$
        \line(1,0){245}} \end{flushright} \narrowtext \noindent}

 \input{psfig}
\draft{}
\begin{document} 
 
\title 
{ Magneto-optical study of magnetic flux penetration 
into a current-carrying  high-temperature 
superconductor strip} 

\author{ 
        M. E. Gaevski, A. V. Bobyl, and D. V. Shantsev}
\address{ 
        A. F. Ioffe Physico-Technical Institute, Polytechnicheskaya 26, 
        St.Petersburg 194021, Russia} 
\author{Y. M. Galperin,\thanks{Also with 
A. F. Ioffe Physico-Technical Institute}
        T. H. Johansen,\thanks{E-mail:
t.h.johansen@fys.uio.no}  M. Baziljevich, and H. Bratsberg}
\address{Department of Physics, University of Oslo, P. O. Box 1048
        Blindern, 0316 Oslo, Norway}
\author{S. F. Karmanenko}
\address{Electro-Technical University, St. Petersburg 197376, Russia}

\date{Received \today} 
\maketitle 

\begin{abstract} 
Magnetic flux distribution across a high-temperature superconductor
strip is measured using magneto-optical imaging at 15~K. Both the
current-carrying state and the remanent state after transport current
are studied up to the currents $0.97 I_c$ where $I_c$ is the critical
current. To avoid  overheating of the sample current pulses with
duration 50 ms were employed. The results are compared with
predictions of the Bean model for the thin strip geometry.  In the
current-carrying state, reasonable agreement is found. However, there
is a systematic deviation -- the flux  penetration is deeper than
theoretically predicted.  A much better agreement is  achieved   by
accounting for flux creep as shown by our computer simulations.  In
the remanent state, the Bean model fails to explain the experimental
results. The results for the currents $I \le 0.7 I_c$ can be
understood within  the framework of our flux creep
simulations. However, after the currents  $I > 0.7 I_c$ the total flux
trapped in a strip is substantially less than  predicted by the
simulations. Furthermore, it {\em decreases}  with {\em increasing}
current.  Excessive dissipation of power in the annihilation zone
formed in the remanent state  is believed to be the source of this
unexpected behavior.  
\end{abstract} 

\pacs{PACS numbers: 74.76.Bz,74.60.Ge,74.60.Jg,78.20Ls}
 
\narrowtext
\section{Introduction} 
 
Spatially resolved studies  of magnetic flux penetration into
high-temperature superconductor (HTSC) films are extensively
performed during the last few years.
Modern experimental techniques, in particular Hall micro-probe
measurements and magneto-optical (MO) techniques, allow   
local magnetic field distributions in various HTSC structures to be
investigated with rather high spatial resolution. As a result, a
quantitative comparison between experimental flux density profiles and
theoretical 
predictions has become possible. Most of the comparisons are done    
within the framework of the critical state model (CSM)\cite{Bean}.  
Based on this model theoretical calculations of the field 
distributions for many practical geometries are carried out. In
particular, the flux profiles in an infinite thin strip placed in a
perpendicular magnetic field, or carrying a transport current are calculated 
in Refs.~\onlinecite{norris,BrIn,zeld1}.  
Most experimental  studies of flux penetration, see
e. g. Refs.~\onlinecite{joh1,shuster,greissen,polyanskii},  are 
focused on the behavior  of samples placed in an external magnetic
field, as well as on   the remanent state after the field is switched-off.
The results appear in a good agreement with experiment.

Meanwhile, we are aware of only a few papers devoted to  experimental
investigation of the self-field of transport
currents~\cite{vv1,ind1,sh1,john1,welp,pash,oota,herrmann}.
Unfortunately, these investigations do 
not allow a simple comparison to the theory. Indeed, some of
them\cite{vv1,john1,welp,pash,oota,herrmann} present results for
samples of rather complicated 
geometry, e. g., for tapes.
Others\cite{ind1,sh1} report results for currents much less than the
critical current, $I_c$.
 
The aim of this work is to study by MO imaging the flux penetration
into a strip with transport current. It seems most interesting to have
the current $I$ as close as possible to the critical one. To reach
this aim one needs narrow strips where the critical current is not too
large to be carried by contacts without their destruction. On the other
hand, the wider the  strip the better the relative spatial resolution
of MO imaging. By optimizing both the strip's width and other
experimental conditions  we managed to obtain flux profiles both in
the current-carrying and in the remanent state with a resolution
sufficient for comparison with theory.
 
In Sec.~\ref{sam} the samples and the experimental technique are
described.  The results for flux profiles and reconstructed current
distributions  are compared to the CSM in Sec.~\ref{results}. It is
shown that the deviations are fairly small   in the current-carrying
state. However, they are pronounced in the  remanent state after
applied current. The main deviation is a deeper  flux penetration
inside the strip compared to predictions of the CSM.  To understand
the source of the  deviation we have carried out numerical simulations
of the field and  current profiles taking into account flux creep. The
results of these simulations are discussed in Sec.~IV. Creep appears
to explain all the experimental results for the current-carrying
state, as well as the results for the remanent state after relatively
small currents, $I \lesssim 0.7\,I_c$. In the remanent  state after
large currents the experimental profiles could not be  explained
either by CSM or by flux creep. It seems that thermal effects are
responsible for this behavior.

\section{Experimental} \label{sam}

\subsection{Sample preparation}

Films of YBa$_2$Cu$_3$O$_{7-\delta}$ were grown by dc magnetron
sputtering  \cite{kar1} on LaAlO$_3$ substrates.  X-ray and Raman
spectroscopy analysis confirmed that the films were  $c$-axis oriented
and of a high structural perfection. Several bridges  were formed from
each film by  a standard lithography procedure. Their dimensions are
$500\times 110\times 0.2$ $\mu$m$^3$.  To minimize temperature
increase caused by  Joule heating  the contact pads were made as wide
as possible.  They are displaced to the side of the structure allowing
the MO indicator film to be placed as close to  the bridge as
possible.  To provide low contact resistance they were covered with a
Ag layer, and Au  wires of 50 $\mu$m diameter were attached by thermal
compression. The boundary resistance of Ag/YBa$_2$Cu$_3$O$_{7-\delta}$
interface was as low as 10$^{-3}-10^{-4}\,\Omega\,$cm$^{2}$,
while the area of one contact pad was $\approx 0.25$~cm$^2$.
The bottom of 500 micron substrate was held at a constant
temperature. Since the thermal conductivity of LaAlO$_3$
is\cite{LaAlO}  0.1~W/(cm \, K), the temperature rise at the
interface induced 
by Joule heating due to currents up to 6~A is always
less than 0.1~K. Thus, resulting heating of the YBa$_2$Cu$_3$O$_{7-\delta}$
bridge situated 500 micron away from the contact pads
is negligible.

An initial selection of the structures having the smoothest surface,
i.e., height of over-growth less than 3 $\mu$m, was made using SEM.
Next, bridges with pronounced weak links, $T_c$-inhomogeneities and
other defects which reduce the total critical current $I_c$ were
eliminated by means of low-temperature SEM,\cite{LTSEM} in addition to
current-voltage measurements made by a standard four-probe  scheme.
As a result, the bridges used for the final investigations had a
critical current density, $j_c$, larger than  $10^6$~A/cm$^2$ at
77~K.

\subsection{Magneto-optical imaging}

Our flux visualization system is based on the Faraday rotation of  a
polarized light beam illuminating an MO-active indicator film that  we
place directly on top of the sample's surface. The rotation angle
increases with the magnitude of the local magnetic field perpendicular
to  the HTSC film. By using crossed polarizers in an optical
microscope one can directly visualize and quantify the field
distribution across the sample area. As Faraday-active indicator we
use a  Bi-doped yttrium iron garnet film with in-plane
anisotropy\cite{dor1}. The indicator film was deposited to a thickness
of 5 $\mu$m by liquid phase epitaxy on a gadolinium gallium garnet
substrate. Finally, a thin layer of aluminum was evaporated onto the
film in order to reflect the incident light and thus providing a
double Faraday rotation of the light beam.   The images were recorded
with an eight-bit Kodak DCS 420 CCD camera  and transferred to a
computer for processing. The conversion the gray level of the image into
magnetic field 
values is based on a careful calibration of Bi:YIG indicator
response to a range of controlled perpendicular magnetic field
as seen by the CCD camera through the microscope (see also
Ref. \onlinecite{joh1}).
After each series of measurement, the
temperature was increased above $T_c$ and the in-situ calibration of
the indicator film was carried out.
As a result, possible errors caused by inhomogeneities of both
indicator film and light intensity were excluded.

\begin{figure}  
\centerline{ \psfig{figure=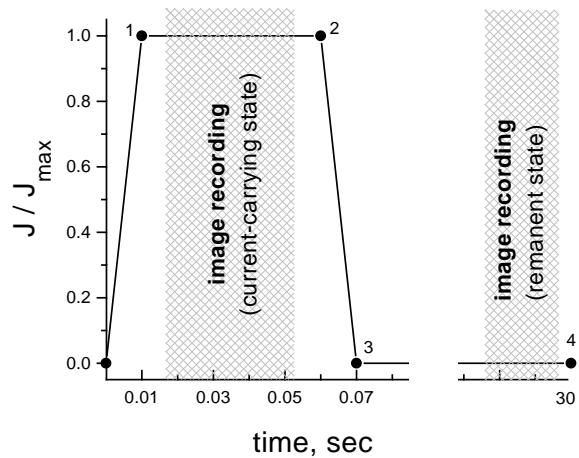,width=8cm}} 
\caption{Temporal profile of the pulsed transport current. The exact
time  intervals of the MO image recording in the current-carrying
state  and the subsequent remanent state are indicated.   } 
\label{f_time} 
\end{figure}

To avoid overheating of the HTSC bridges as the current approaches $I_c$
a specific experimental procedure was developed.
Short current
pulses were applied and synchronized with the
camera recording as shown in Fig.~\ref{f_time}.
\noindent
The transport current with a rise time of 10~ms was applied 20-30~ms
before an image was recorded. The exposure time was 35~ms, after which
the current was ramped to zero in 10~ms. After
waiting another 20-30 seconds
the remanent field distribution was measured.

\begin{figure} 
\centerline{
\psfig{figure=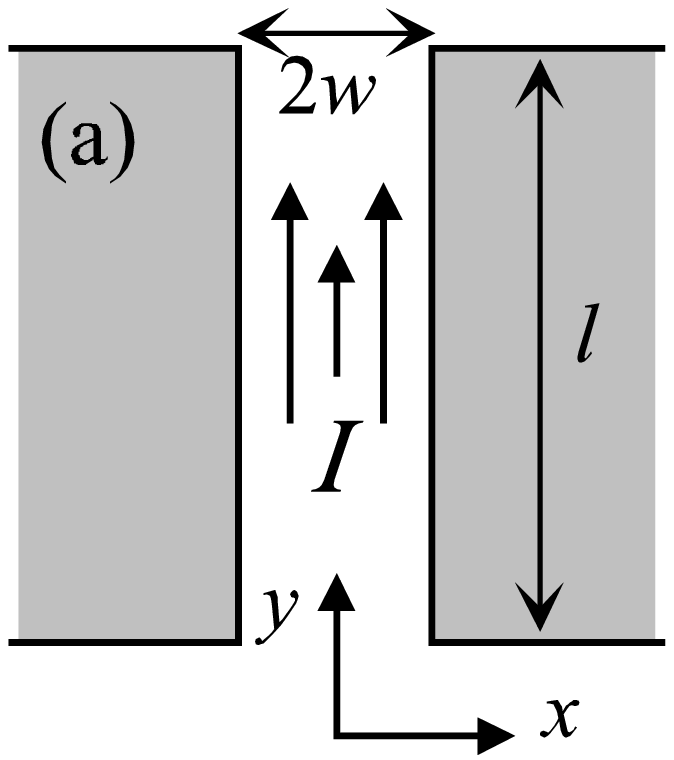,width=6.4cm}
}
\centerline{
\psfig{figure=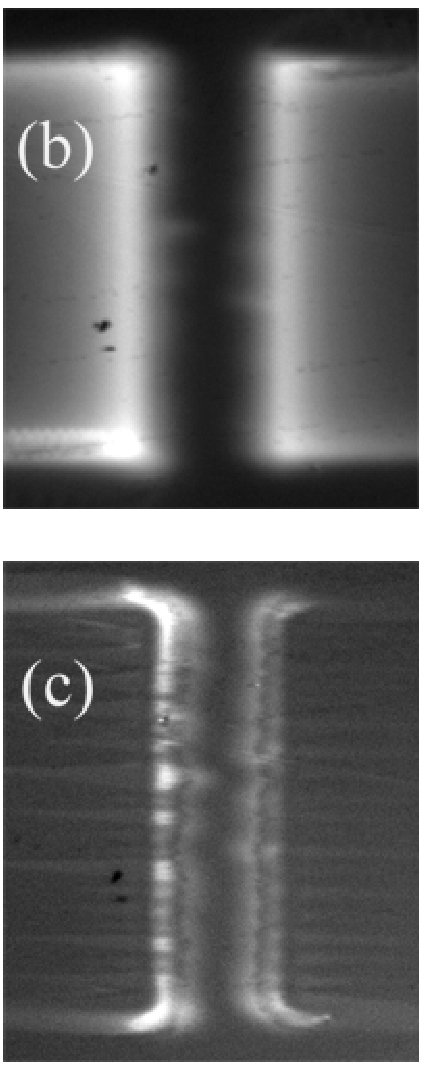,width=4.9cm}
}
\vspace{3mm}
\caption{(a) Sketch of the studied YBa$_2$Cu$_3$O$_{7-\delta}$ bridge
structure. (b) MO image of flux distribution with the bridge in the
current-carrying state   with $I=4.16$~A.  (c) MO image of flux
distribution with the bridge in the remanent state  after the
applied-current state shown in (b).  Two inner bright lines  near the
sample edges represent the original trapped flux, the outer bright
lines represent penetration of the 
oppositely directed return field, while dark lines between them
correspond to  regions of 
vortex-antivortex annihilation.  Strong contrast enhancement was
applied to the image in (c).  }
\label{f_scheme}
\end{figure}

\section{Results and comparison with CSM} \label{results}

Throughout the paper we use the following notations, see
Fig.~\ref{f_scheme}.  The $x$-axis is directed across the bridge, the
edges being located at  $x=\pm w$. The $y$-axis points along the
bridge,  and the $z$-axis is normal to the film plane.  The
$z$-component of the flux density is denoted by  $B$.   The profiles
$|B(x)|$ were always measured for a fixed $y$ in the central part of
the strip minimizing   the effect of the stray field of the contact
pads.  Although the MO measurements give the distribution  of
$|B(x)|$, we could in the simple geometry under consideration  always
determine the sign of $B$  by inspection of the images.  We let $J(x)$
denote the sheet current density defined as $J(x)=\int{j(x,z)dz}$,
where $j(x,z)$ is the    current density. For brevity, we will   often
use current density also for $J(x)$. As the bridge thickness
is much less than its width, the theoretical results for
the thin strip geometry\cite{BrIn,zeld1} are used below. In our
experiments, the bridge thickness is also of the order of the
penetration depth $\lambda$, hence its magnetic properties are fully
characterized by the two-dimensional flux distribution at the surface.

\subsection{Current-carrying bridge}

\begin{figure}
\centerline{ \psfig{figure=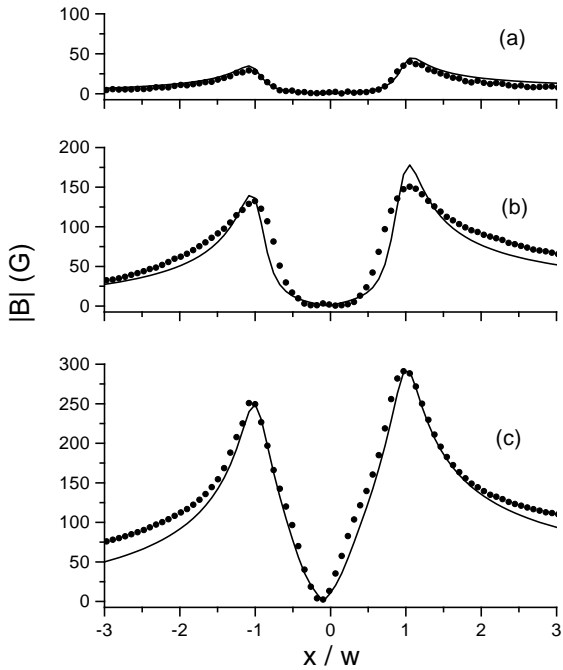,width=8cm}}
\caption{  Profiles of the perpendicular magnetic field produced at a
distance $h=8 \ \mu$m    above the bridge carrying transport currents.
Symbols show the   experimental data (only 1/5 of all measured points
are shown),  and the solid lines represent  CSM calculations from the
\protect{Eqs.(\ref{JJtHa}) and (\ref{Bh})}. $I/I_c=0.13$(a), 0.53(b),
0.97(c), where $I_c=5.9$~A.   } \label{f_BJt}
\end{figure}

Flux density distributions for a strip carrying a transport  current
$I$ were measured for currents up to $I$=5.72~A.  The MO image for the
current $I=4.16$~A is shown in Fig.~\ref{f_scheme}(b).  Three profiles of
the flux density $|B(x)|$ taken across the strip   are shown in
Fig.~\ref{f_BJt}.  The profiles have maxima near the strip edges  and
a minimum in between.  Actually, the left and the right parts of the
profiles correspond  to induction of opposite sign.  As the current
increases, (a)-(c), the magnetic flux  penetrates deeper and the flux
free Meissner region in the center of the bridge decreases in size.

The experimental MO data are interpreted in the framework of the
CSM. As discussed in Ref.~\onlinecite{joh1},  one should account for
the finite distance between the  sample and the MO indicator film.
The perpendicular magnetic field, $B$, at the height $h$ above the
center of the bridge  can be calculated from the current density
distribution $J(x)$  as\cite{joh1} 
\begin{equation} \label{Bh} 
B(x) = \frac{\mu_0}{2\pi} \int_{-w}^{w}
        { \frac{x'-x}{h^2+ \left( x'-x \right)^2 } \, J(x')\, dx'\,+B_a. } 
\end{equation} 
Here $B_a$ is the external magnetic induction.  
The current density distribution in a strip carrying a transport 
current $I$ can be written for the Bean model as\cite{BrIn,zeld1} 
\begin{equation} 
\displaystyle{\frac{J(x)}{J_c}} = \left\{ \begin{array}{lr} 
              \displaystyle{\frac{2}{\pi}} \arctan \left( 
                 \sqrt{\displaystyle{\frac{w^2-a^2}{a^2-x^2}}} 
              \right), & |x|<a \\ 
              1, & a<|x|<w 
              \end{array} 
       \right. , 
\label{JJt} 
\end{equation} 
where $a=w\sqrt{1-(I/I_c)^2}$, and $I_c=2w\, J_c$ is the critical
current.  As Eqs.~(\ref{Bh}) and (\ref{JJt}) yield a symmetric
$|B(x)|$-profile   we find it necessary to account also for the stray
field of the   contact pads in order to reproduce the slight asymmetry
in the $|B(x)|$ data.  The current in the pads produces   near the
central part of the bridge  a magnetic field which acts as an
additional external field  varying slowly in space. This allows us to
employ the results of the CSM for the case of a transport  current
superimposed by a weak external magnetic field.\cite{end1}  According
to this theory\cite{BrIn,zeld1},   the current density distribution
Eq.~(\ref{JJt})  is modified  to yield $J(x)=J_c$ at  $x$ within the
intervals $(-w,p-a)$,  $(p+a,w)$ and 
\begin{eqnarray} 
\displaystyle{\frac{J(x)}{J_c}} &=& 
              \displaystyle{\frac{1}{\pi} }\left( 
                 \arcsin\displaystyle{ \frac{(x-p)(w-p)-a^2}{a(w-x)}} 
\right.\nonumber \\
&& \left. -                  \arcsin 
              \displaystyle{\frac{(x-p)(w+p)+a^2}{a(w+x)}}\right) + 1 
              \label{JJtHa} 
\end{eqnarray} 
for $p-a < x < p+a$. Here 
 
\begin{eqnarray*}
a&=&\frac{w}{\cosh \left(B_a/B_c \right)} 
   \sqrt{ 1 - \left( \frac{I}{I_c} \right)^2 }, \\
p&=w& \frac{I}{I_c} \tanh\left( \frac{ B_a}{B_c} \right) 
\, , \quad
B_c \equiv \frac{\mu_0 J_c}{\pi}\, . 
\end{eqnarray*}

Profiles of $|B(x)|$ calculated from Eqs.~(\ref{JJtHa}) and (\ref{Bh})
are shown    in Fig.~\ref{f_BJt} by the solid lines.  Here $I_c$ and
$h$ are parameters determined by fitting the  CSM profiles to
experimental data. The values $I_c=5.9$~A, $h= 8 \  \mu$m yield the
best fit for the profiles measured at all currents $I$. Note that
$I_c$ has a physical meaning of the critical current for CSM model, at
which the  magnetic field fully penetrates the bridge.  
 
One can notice in such experiments, and e. g. in Fig.~\ref{f_BJt}(b),
that the experimental  penetration of the magnetic field is deeper
than the CSM prediction.   This characteristic deviation can be
accounted for by introducing flux creep,  as  will be discussed in
detail  in the next section. 
  
With $h$ being a known parameter one can also  directly determine the
current distributions  across the strip, $J(x)$, from the experimental
$B(x)$-profiles.   This can be done on a model-independent basis
according to an inversion scheme developed in Ref.~\onlinecite{joh1}.
Since the  accuracy of the inversion is limited by the distance $h$,
we chose a  step size along the $x$-axis close to $\Delta = h/2$.
Specifying the coordinates in units of $\Delta$, i.e.,  $x=n \Delta, \
x'=n' \Delta, \ h= d \Delta$,  and applying a Hannig window filtering
function one obtains\cite{joh1}: 
\begin{eqnarray} 
&&J(n)=\sum_{n'} \frac{n-n'}{\mu_0 \pi} \left( 
 \frac {1-(-1)^{n-n'}e^{\pi d}} {d^2+(n-n')^2}\right. \nonumber \\ &&\left.
 \ + 
 \frac {\left[ d^2+(n-n')^2-1 \right] \left[ 1-(-1)^{n-n'}e^{\pi d}\right]} 
       {\left[ d^2+(n-n'+1)^2 \right] \left[ d^2+(n-n'-1)^2 \right]} 
       \right) B(n'). 
\label{jb} 
\end{eqnarray} 

\begin{figure}
\centerline{ \psfig{figure=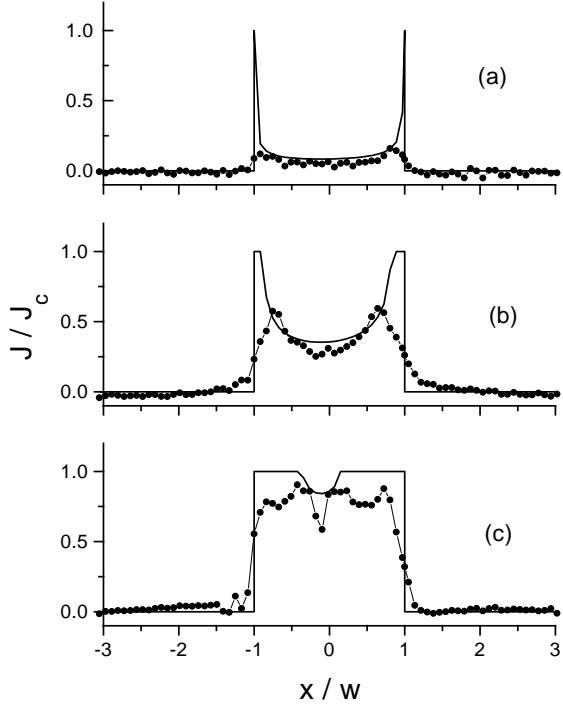,width=8cm}}
\caption{  Distributions of current density in the current-carrying
strip. Symbols show the distributions inferred from experimental data
using Eq.~(\protect{\ref{jb}}).  The predictions of the CSM,
Eq.~(\protect{\ref{JJtHa}}) are shown  as full lines.
$I/I_c=0.13$(a), 0.53(b), 0.97(c), where $I_c=5.9$~A.  } \label{f_JJt}
\end{figure}

The current density profiles calculated from  experimental $B(n)$-data
using this formula are shown in Fig.~\ref{f_JJt}.  The figure also
shows the results obtained from Eq.~(\ref{JJtHa}) using  the value of
$I_c$ determined by the fitting of  $|B(x)|$-profiles. As seen from
Fig.~\ref{f_JJt}, the experimental  profiles trace all qualitative
features of the theoretical curves.  In particular, the minimum in the
current density   near the bridge center predicted by the CSM can
clearly be distinguished  in all experimental curves. Moreover, the
position  of the minimum is found to be slightly shifted towards
negative $x$, a behavior in full agreement with the theory   when the
effect of contact   pad stray fields is accounted for. 
 
As first pointed out in Ref.~\onlinecite{joh1}, the MO YIG indicators
with in-plane anisotropy respond not only to $B_z$  but also to the
component $B_x$ parallel  to the film. In the data presented we always
made the proper corrections   according to the method  suggested  in
Ref.~\onlinecite{joh1}, i.e., $B_x (x)$ component was  calculated from
the current profiles by Biot-Savart law and then  used to re-normalize
the profile $B_z(x)$.  This procedure is  converging very rapidly. 

To check self-consistency of our inversion calculations we integrated
the  current density over $x$. Indeed, The total current was  always
equal to the transport current  passed through the bridge   within an
accuracy of $5\%$.   

\noindent  
Some deviations between experimental and
theoretical current profile slopes  are seen near the sample edges
where the theoretical profiles  are discontinuous. The observed
smearing of the current profiles reconstructed  from the induction
data can be ascribed mainly to discreteness of the points used  in the
inverse  calculation, and possibly a slight structural imperfections
near the very edge of the sample.  
 
\subsection{Bridge in a remanent state after transport current} 
 
When the transport current $I$ is switched-off, the magnetic flux  in
the inner part of the strip remains trapped.  The return field of this
trapped flux will re-magnetize the edge regions of  the strip and the
flux of opposite sign penetrates an outer rim.   As a result,  the
measured remanent distribution $|B(x)|$  should display two peaks in
each half of the bridge; one   representing the maximum trapped flux,
and another near the  edge indicating the maxima in the reverse flux.
In this oscillatory flux profile  typical values of $|B_z|$ are
significantly  lower than in the current-carrying state. Thus,  MO
studies become substantially more  difficult in the remanent state,
and flux profiles with  reasonable signal-to-noise ratio were recorded
only after relatively large transport currents, $I>0.6 I_c=3.68$~A.

\begin{figure}
\centerline{ \psfig{figure=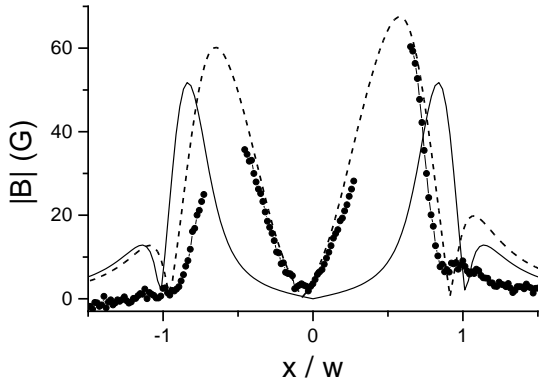,width=8cm}}
\caption{  Remanent profiles of the absolute value of perpendicular
magnetic field  at the distance  $h=8 \ \mu$m above the bridge  after
applying transport current $I=4.16$~A.  Dots -- experiment, solid line
-- CSM, Eqs.~(\protect{\ref{JRJt}}),~(\protect{\ref{Bh}}) with
$I_c=5.9$~A, dashed line --   flux creep simulations.   A large amount
of flux trapped near the bridge center is in strong  contradiction
with CSM and can be explained by flux creep.  Experimental points in
the regions where the parallel component $B_x(x)$ changes sign and the
response of indicator film is suppressed have been removed from the
plot.   } \label{f_BRJt} 
\end{figure}

The MO image of flux density distribution in the remanent state  after
 switching off a transport current of 4.16 A is shown  in
 Fig.~\ref{f_scheme}(c).  Figure~\ref{f_BRJt} shows the flux density
 profile taken across   the strip in this remanent state.  It can be
 seen that there are large maxima of  trapped flux in the center of
 the bridge.  Also, the weaker maxima of reverse flux are visible near
 the edges.   In the figure we have removed the experimental points
 corresponding to the regions where   the MO image is governed by
 disturbing zig-zag domain pattern in the indicator  film. This occurs
 where $B_x$ changes sign.

The current distribution  derived from the Bean model for the remanent
state is\cite{BrIn,zeld1}  

\wide
\begin{equation} 
\displaystyle{\frac{J(x)}{J_c}} = \left\{ \begin{array}{lc} 
              \displaystyle{\frac{2}{\pi}} \left[ 
              \arctan 
              \left(\sqrt{\displaystyle{\frac{w^2-a^2}{a^2-x^2}}} 
              \right) - 
              2 \arctan 
              \left(\sqrt{\displaystyle{\frac{w^2-b^2}{b^2-x^2}}} 
              \right) 
              \right] , & -a<x<a, \\ 
               1- \displaystyle{\frac{4}{\pi}} 
              \arctan 
              \left(\sqrt{\displaystyle{\frac{w^2-b^2}{b^2-x^2}}} 
              \right) , & a<|x|<b 
              \end{array} 
       \right.\, , 
\label{JRJt} 
\end{equation} 
\narrow 
 where $a=w\sqrt{1-(I/I_c)^2}, \  b=w\sqrt{1-(I/2I_c)^2}$ and $I$ is
the maximal current. At $b < |x| <w $ the   current density is equal
to $-J_c$.  The small contribution from  the field generated by the
contact remanent currents  is here neglected.  

The magnetic field profiles calculated using Eqs.~(\ref{JRJt}) and
(\ref{Bh})  are shown in Fig.~\ref{f_BRJt}  (solid line) together with
the experimental data.  We used the same values for  $I_c$ and   $h$
as obtained by fitting results for current-carrying state.  Evidently,
there is here a significant  deviation between our data and the CSM
description.  The main deviation is that the trapped flux maxima  in
the experimental curve are shifted  towards the center. Again, this
can be shown to be an effect of flux creep  as discussed in the next
section.   Another deviation seen in Fig.~\ref{f_BRJt} is that the
maxima near the edges   are hardly visible experimentally.   This is
most likely due to the nonlinear response   of the optical detection
system which leads to reduced sensitivity  at low Faraday rotation
angles, i.e., at small induction values.  Another source of smearing
is   that the width of the maxima near the edges is comparable to the
thickness of MO  indicator film. 
 
\begin{figure} 
\centerline{ \psfig{figure=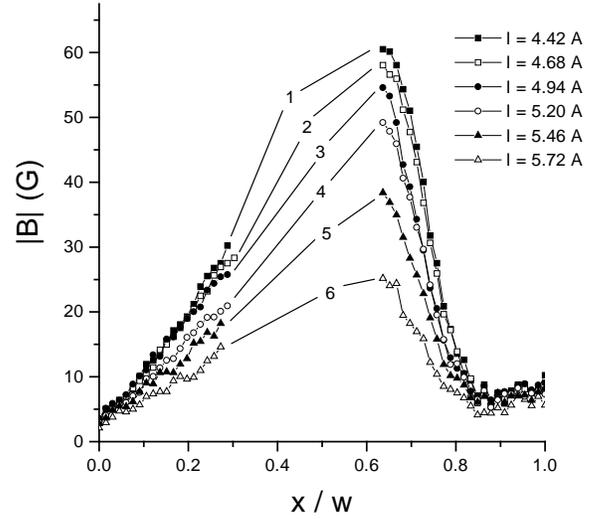,width=8cm}} 
\caption{  Experimental profiles of the absolute value of
perpendicular magnetic field  in the remanent state  after different
transport currents $I$:   (1) - 4.42~A,  (2) - 4.68~A,  (3) - 4.94~A,
(4) - 5.20~A,  (5) - 5.46~A,  (6) - 5.72~A.  Only the right half of
the bridge is shown.  Note the unusual $I$-dependence of the trapped
flux:  the {\em larger} the current the {\em smaller} the density of
trapped flux in the bridge.  } \label{f_6cur} 
\end{figure}

\begin{figure} 
\centerline{ \psfig{figure=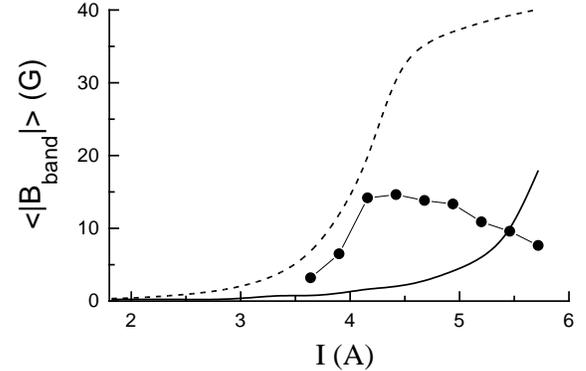,width=8cm}} 
\caption{  Flux trapped within the band $\, -0.4 w <x <0.2 w\,$ in the
remanent state after transport current $I$ as a function of $I$.   The
band corresponds to reliable MO data  shown in Fig.~\ref{f_BRJt}, the
flux in normalized to the band width.   Dots -- experimental
results. Solid and dashed lines --   CSM and flux creep simulations
respectively. For $I \le 4.16$~A the experiment is in a qualitative
agreement with flux creep,   however for larger currents a striking
steady decrease in trapped flux is observed.}  
\label{f_nonmon} 
\end{figure}

The experimental flux profiles in the remanent state   after large
currents $I > 4.16$ show even more intriguing behavior.
Fig.~\ref{f_6cur} shows six  remanent flux profiles across the right
half of the bridge  after applying currents ranging from 4.42~A to
5.72~A.    The observed {\em decrease}  of the trapped flux with {\em
increasing} transport current seems  highly unexpected. Such a
behavior definitely contradicts the CSM.  To make a quantitative
comparison we have integrated the flux trapped  in the band $-0.4 w <x
<0.2 w$ , where the MO data are reliable.   The dependence of the
total flux on the current is shown in Fig.~\ref{f_nonmon}.  At small
currents, $I \le 4.16$ A,  the behavior is  normal  as the amount of
trapped flux increases with the current.  At larger currents, however,
the   trapped flux levels out and even starts to decrease as $I$
approaches $I_c$.  The full line in the same figure indicates the
predictions of the CSM,  which clearly shows a different
behavior. Neither is the non-monotonous behavior seen experimentally
explained  by flux creep.  

Let us sum up the comparison with the CSM for the current-carrying and
remanent states.  The general features of flux distribution in a strip
with  transport current predicted by the CSM are well confirmed by
the experiments. It should be noted that such  a good agreement was
achieved only by taking into account  a finite distance between the
HTSC film and the MO indicator, the 
 
\noindent
additional field  generated by
contact currents, and the influence of parallel  field component on
the properties of YIG indicator. However, a general  trend of a deeper
flux penetration compared to the CSM prediction can be  traced.
In the remanent state  the deviations from the CSM are more pronounced,
and experimental flux 
profiles after currents $I > 0.7I_c$ have
qualitatively different shape. In particular, the flux  trapped inside
the bridge can be several times greater than predicted  and it depends
on the previously applied current in a non-monotonous way. 

Comparing the experimental results with the predictions of CSM we
employed the expressions (\ref{JJtHa}) and (\ref{JRJt}) based on the
Bean model.   Thus, we assumed the critical current density $J_c$ to
be  $B$-independent. We have checked validity of this assumption
studying  penetration of an applied magnetic field into a bridge of
the same  geometry fabricated on the same film. We could trace rather
weak dependence of   $J_c$ versus $B$  only above  $B \approx 200$
G. For smaller $B$  all  the results, including the results for the
remanent state after  external field,  were compatible with the CSM
predictions, as in  Ref.~\onlinecite{joh1}. Consequently, there is no
reason to attribute  the deviations observed in the transport current
regime to a $B$-dependence of $J_c$.  

\section{Flux creep} 
 
The basic assumption of the CSM is that in the regions where the local
current density $J$ is less than the critical one $J_c$, the flux
lines do not move. This assumption does not hold for any finite
temperature  because of thermally-activated flux motion or flux creep.
As a result, a small amount of magnetic flux will penetrate into the
regions  with $J<J_c$.  We suggest that this excess flux penetration
is responsible for   the smoothening of the experimental profiles
$B(x)$ relative to  the  CSM profiles.  
 
To verify this suggestion we have carried out computer simulations  of
flux penetration into a strip in the flux creep regime.  Simulations
of this type  have previously proven to be very powerful in the
analysis  of flux  behavior.  In particular, they have been used   to
analyze magnetization curves\cite{kes,burlachkov},   magnetic
relaxation data\cite{burlachkov,chinese}, and  ac susceptibility for
various geometries\cite{GurBra97}.  Flux creep simulations have also
been used to explain features of  the flux penetration into thin HTSC
samples in applied magnetic field\cite{shuster,brandt97}.  For the
first time we present here creep simulations to analyze  flux dynamics
in a strip with transport current and the subsequent  remanent state.  
 
\subsection{Model} 
 
Motion of magnetic flux is governed by the equation 
\begin{equation} 
  \frac{\partial B}{\partial t} = - 
  \frac{\partial}{\partial x} \left( vB \right), 
\label{dBdt} 
\end{equation} 
where $v$ is the vortex velocity.
The velocity
$v$ is assumed to be dependent on the current density $J$ and the
temperature $T$ as $v=v_0 \exp[-U(J)/kT]$ where $U(J)$ is the
current-dependent activation energy due to vortex pinning.     Its
dependence on the current density is extensively discussed in the
literature, see for a review Refs.~\onlinecite{blatter,yeshurun}. 
 
A conventional approach to the flux creep is based upon the  linear
(Anderson-Kim) relation $U(J)=U_c(1-J/J_{cp})$ for the pinning
energy. Here we use notation $J_{cp}$ for the {\em depinning} critical
current density which is different from the CSM critical current
density, $J_c$. The Anderson-Kim approximation is fairly good at
$J_{cp}-J \ll J_{cp}$. However, due to low pinning energies in HTSCs,
the time dependence of the current density $J$ appears very
pronounced, and during the time of experiment $J$ can reach the values
well below $J_{cp}$. As a result, to describe experimental data one
should take into account a nonlinear character of the $U(J)$
dependence\cite{blatter,yeshurun}.  The  usual way is to express the
pinning energy as a power law  function of the current density, $U(J)
\propto J^{-\mu}$. Such a  dependence follows from several theoretical
models based on the  concept of collective creep, the values of $\mu$
being dependent on  vortex density, the current, the pinning strength,
and  dimensionality\cite{blatter}.  Since we are interested in the
region  of low fields, the value  $\mu=1/7$ seems most appropriate, as
it applies  to the case of a single-vortex creep at  high current
density and low temperature\cite{1/7}.  
We use the expression $U(J)=U_c\left[\left(J_{cp}/J \right)^\mu-1
\right]$ for the activation energy. In such a normalization
$U(J_{cp})=0$ and $J_{cp}$ retains the meaning of a depinning critical
current. The current-independent factor in the vortex velocity equals
$v_0\, \exp(U_c/kT)$   where $v_0$ is a quantity of the order of the
flux-flow velocity. 
   
The numerical integration of equation (\ref{dBdt}) was carried out by
a single step method, similar to the one  reported in
Ref.~\onlinecite{chinese}.   Having the field distribution $B(x,t)$ at
time $t$, we calculate the    corresponding  current density
distribution as\cite{BrIn} 
\begin{eqnarray} 
  J(x,t) &=& \frac{2}{\pi\mu_0} \int^{w}_{-w} \frac{B(x',t)-B_a(t)}{x-x'} 
  \sqrt{\frac{w^2-x'^2}{w^2-x^2}} dx' \nonumber \\ &&  + 
  \frac{I(t)}{\pi \sqrt{w^2-x^2}}\, . 
\end{eqnarray} 
Here $I(t)$ and $B_a(t)$ are the time-dependent total transport
current and applied field, respectively.  The expression follows from
the Maxwell law for the thin film geometry.  Then the quantity
$\partial B (x,t)/\partial t$ is calculated  from
Eq.~(\ref{dBdt}). The new distribution of magnetic field is calculated
as   $B(x,t)+\delta B(x,t)$, where  $\delta B(x,t) = \delta t  \cdot
(\partial B/\partial t) $. Here a  time  increment  $\delta t$   is
chosen  so that    $\delta B(x,t) \le   \delta B_{\max}= 0.0001 \mu_0
J_{cp}$ for any $x$.  Then we come to the next step and so on.

There are two independent parameters in the model -- the product
$v_0\, \exp(U_c/kT)$ and the ratio $ J_{cp}^\mu U_c/\left(kT\right)$.
Unfortunately, it seems very difficult to estimate these parameters
from the literature because of a very large scatter in the  published
values of the pinning energy $U_c$   for
YBa$_2$Cu$_3$O$_{7-\delta}$. We determine the ratio   $J_{cp}^\mu
U_c/\left(kT\right)$ by fitting the  experimental flux profiles and
estimate the quantity $v_0\, \exp(U_c/kT)$
from a separate  experiment on magnetic relaxation.

\begin{figure} 
\centerline{ \psfig{figure=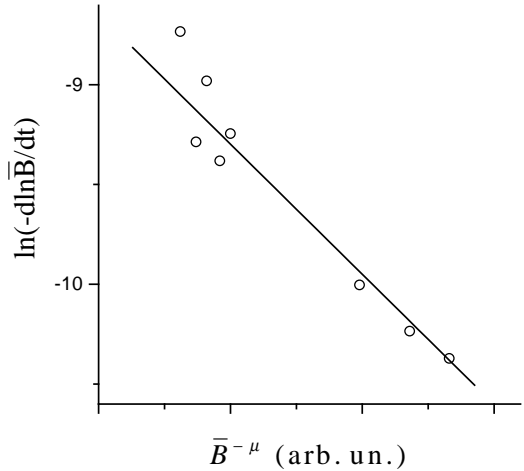,width=8cm}} 
\caption{  On the determination of the flux creep parameters from the
experimental MO data on magnetic  relaxation. 
${\bar B} (t)$ is the remanent field in the central region of the
bridge after switching off the external field 100~mT. 
The experimental 
dependence $\ln (- d\, \ln {\bar B}/dt)$ versus ${\bar B}^\mu$  with $\mu=1/7$
is fitted by a straight line according to Eq.~(\protect{\ref{relax}}).
} \label{f_relax} 
\end{figure} 

In the relaxation experiment the sample was cooled down to 15~K in
external magnetic  field 100~mT, which is about $4 B_c$ for
our bridge. This high field ensures that the remanent state is fully
penetrated by the current\cite{zeld1,BrIn}. Then the field was
switched off and
the time dependence of the  remanent field was measured
by MO technique in the time window 10$^1$-10$^3$~s. From now on we
focus on the peak field value $\overline{B}(t)$ observed in the
central region of the bridge and averaged along its length.
Instead of presenting of the relaxation of $\overline{B}$ directly, we
chose to plot the logarithmic derivative $\ln (\partial \ln
\overline{B}/\partial t)$ versus $\overline{B}^{\mu}$. The reason for this type
of plot is the following.
 As was shown in
Ref.~\onlinecite{burlachkov}, the pinning energy  $U(x)$ remains
almost the same locally at any stage in a  relaxation process. That is a
feature of a  self-organized behavior of magnetic flux which is a
consequence of the  exponential dependence $v \propto \exp[-U(J)/kT]$.
Since $U$ is almost constant $v$ varies also slowly in space, and therefore

\[
\frac{\partial B}{\partial t} \approx v\frac{\partial B}{\partial x}\,
.
\]
We can roughly estimate $\partial B / \partial x $ as
$\overline{B}/w$. Moreover, in a fully penetrated state $\partial B
/\partial x \propto J$, and therefore, $J \propto \overline{B}$.
This approximation together with Eq.~(\ref{dBdt}) yields
\begin{equation}
\ln \left( -  \frac{\partial \ln \overline{B}}{\partial t} \right) \approx
\ln \left[\frac{ v_0}{w}\exp \left(\frac{U_c}{kT}\right)\right]  -
\frac{\text{const}} {\overline{B}^\mu}.
\label{relax}
\end{equation}
The experimental quantity
$\ln \left( - \partial \ln \overline{B} /\partial t
\right)$ is plotted as a function of  $\overline{B}^{-\mu}$ in
Fig.~\ref{f_relax}.
  The experimental points show a nearly
linear dependence, and the line representing
the best fit is  also shown.  From the fit  we
obtain  $v_0 \, \exp (U_c/kT) \approx  10^{24}$ m/s.  Assuming
$v_0$=10 m/s (see, e.g., Ref.~\onlinecite{kes})   we estimate the
pinning energy as  $U_c=(50-55)\, kT \approx 0.08$~meV for $T=15$~K
which is a quite reasonable  value\cite{Hagen}.  The other free
parameter, the sheet current density $J_{cp}$,  was chosen to provide
the best agreement  of
results of flux creep simulations  with experimental profiles
shown in Fig.~\ref{f_BJt}.
It is interesting
to compare  the chosen value $J_{cp} =1.45 \times 10^3 $ A/cm with the
critical sheet current density $J_c$  
determined by  fitting the same experimental data to the CSM.   We
found that $J_{c} \approx 0.37 J_{cp}$. For this current  density,
the   effective barrier $U_{\text{eff}}=U_c\left[ \left(J_{cp}/J_c
\right)^\mu -1 \right]$   appears substantially lower than $U_c$,
about $0.15 \, U_c \approx 8  \, kT$.  It is worth noting that
variation of $\mu$ in the range 1/7-1/3   has minor effect both on the
calculated profiles $B(x)$ and on the relation   $U_{\text{eff}}
\approx 8 \,  kT$.

\subsection{Results}  
 
\begin{figure} 
\centerline{ \psfig{figure=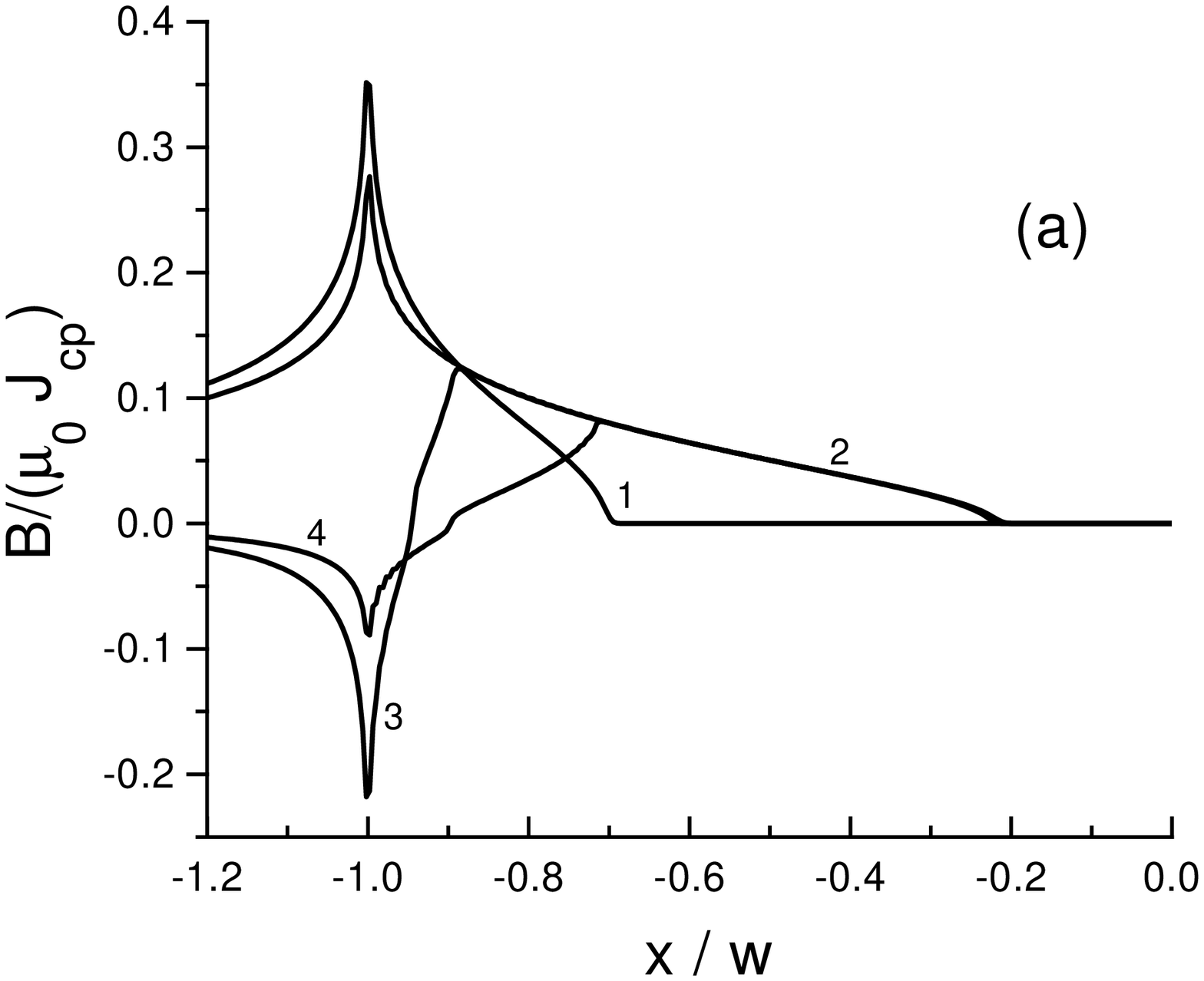,width=8cm}
\psfig{figure=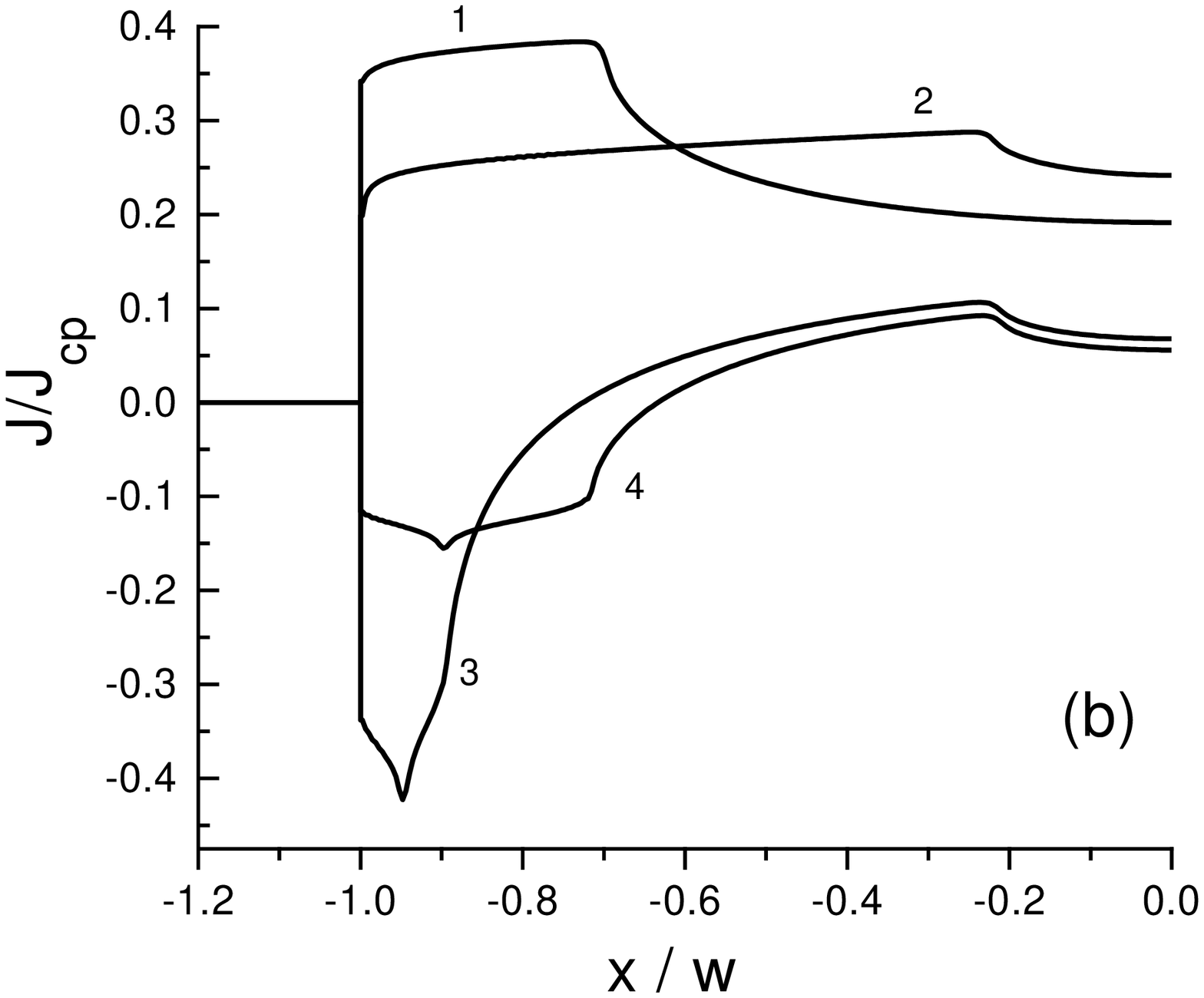,width=8cm}} 
\caption{  Results of flux creep simulation for $J/J_{cp}=0.26$ A.
Evolution of profiles of the magnetic field (a)  and the current
density (b) in the strip plane for the time variation  of the
transport current shown in Fig.~\protect{\ref{f_time}}.  Profiles (a)
and (b) correspond to  different times marked as different points in
Fig.~\protect{\ref{f_time}}.  } \label{f_evol} 
\end{figure}

Let us first discuss general results of flux creep simulations.  Time
evolution of the profiles of current density  and magnetic field
under applying and switching off  a transport current with density
$J=0.26\, J_{cp}$ is shown in Fig.~\ref{f_evol}.  The time dependence
of the transport current was chosen as shown  in Fig.~\ref{f_time} in
accordance with the experimental procedure.  For simplicity, only the
case of zero external field is considered.  Since all the
distributions are symmetric in this case, we discuss  the profiles for
one half of the bridge.

The various curves in Figs.~\ref{f_evol}(a,b) correspond to different
times as marked in Fig.~\ref{f_time}.  The profiles (1-4)
corresponding to current-carrying states  are  very similar to the
ones expected from the CSM.  It is interesting to note that though the
transport current  did not change from the instant 1 to 2, the curves
1 and 2 in  Fig.~\ref{f_evol}b  differ substantially. Though both can
be described by CSM-profiles, they  correspond to different values of
$J_c$. In particular, the transition  from the curve 1 to 2
corresponds to $J_c$ decreasing in  time. Time dependence of effective
$J_c$ can be   seen also from the curves 3 and 4 for the remanent
state  after current.    A possibility to interpret experimental time
evolution of magnetic  flux profiles by  the CSM with time-dependent
$J_c$ has been  previously demonstrated in Ref.~\onlinecite{mcelf}.   
An additional feature of calculated profiles in the remanent state is
a peak of ``negative" current density located  not far from the bridge
edge (cf. with Ref.~\onlinecite{burlachkov}) -- see  curves 3 and 4 on
Fig.~\ref{f_evol}b. The peak's position  corresponds to the
annihilation zone where $B(x)$ changes its sign.  This peak is a
direct consequence of continuity%
\widetext
\narrow

\noindent 
of the flux flow.  Indeed, in the
region of low density of flux lines their velocity  (determined by the
ratio $J/J_{cp}$) should be large to keep the  flux flow continuous.

\begin{figure} 
\centerline{ \psfig{figure=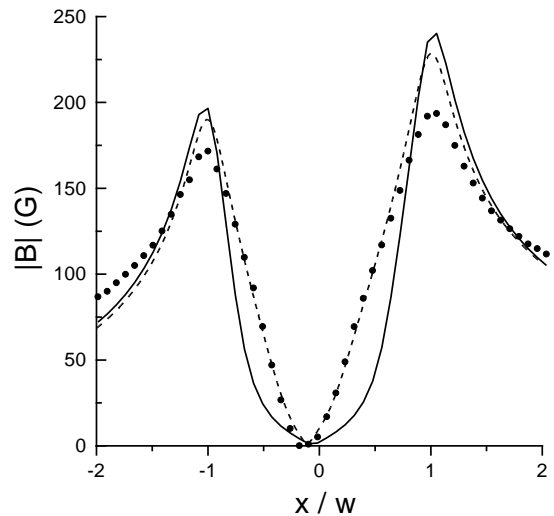,width=8cm}} 
\caption{   Profiles of the absolute value of perpendicular magnetic
field  induced by transport current $I=0.75 I_c$  at the distance $h=8
\ \mu$m above the sample.  Dots --  experiment,  solid line --
calculations along CSM,  \protect{Eqs.~(\ref{JJtHa}),(\ref{Bh})},
dashed line -- results of flux creep simulation. It is
clearly seen that the flux creep  leads to a deeper penetration of
flux comparing to the CSM, in agreement  with experiment.}
\label{f_BJtcreep}
\end{figure}

The dashed line in Fig.~\ref{f_BJtcreep} is a result of our
simulations of the  magnetic field profile for  $I=0.75\,I_c$ at
$t=35$~ms after  switching on the  current. This time delay
corresponds to the center of the plateau in  the time dependence of
the transport current, see Fig.~\ref{f_time},  so it was the  time
when the MO images were recorded. The profile was calculated from  the
simulated current distribution using Eq.~(\ref{Bh}). In a similar  way
the flux profile for the remanent state ($t=30$~s) was
calculated. The result is shown  Fig.~\ref{f_BRJt}.  It is clear that
flux creep provides deeper penetration of the flux into  the inner
regions in agreement with the experiment.  
 
For the  current-carrying state, Fig.~\ref{f_BJtcreep}, the agreement
is fairly  good except two minor discrepancies.  First, the
experimental peaks at the  bridge edges are less than the ones
predicted both by the CSM and by  our simulations. This is a rather
general feature (cf. with Ref.~\onlinecite{joh1}) which is probably
originated from the finite thickness of the indicator film (in our
case 5 $\mu$m), as well as from imperfection of the edges.  Another
discrepancy observed outside the  bridge, is obviously related to
contact  currents. Indeed, Eq.~(\ref{JJtHa}) is based upon the
assumption that the   external field is small\cite{end1} and
homogeneous.   However,  the field generated by the current in the
contacts is actually   inhomogeneous in $x$ direction, and far from
the bridge it is  significantly different from the one in its central
part.  
 
For the remanent states after current $I$, the account of flux creep
provides good agreement with the experiments for $I \le 4.16$ A,
Fig.~\ref{f_BRJt}.  However, at larger currents the experimental
behavior of the trapped flux is qualitatively different. Indeed, both
critical state and flux creep models predict monotonous dependence of
the trapped flux on the current. Non-monotonous  experimental
dependence can serve as an indirect indication that some
non-equilibrium process is responsible for the   flux distributions
observed after large currents.

We believe that formation of the remanent state after very large currents is
strongly influenced by heating effects. As well-known, energy
dissipation due to vortex motion facilitates more
intensive motion. As a 
result, macroscopic  avalanche-like flux  redistributions (flux
jumps)\cite{gur1,McHenry,McHenry2,muller,Mints} can take place. 
Unfortunately, the works we are aware of are focused on  the
 flux jumps in the case of applied magnetic field and slab geometry.
We believe that the case of a remanent state in a strip after
transport current requires a special theoretical treatment. 
Furthermore, the flux motion obviously takes place through
narrow channels  of weak pinning. Consequently, the energy dissipation
 is substantially inhomogeneous that should be taken into account in
 the estimates of local temperature. We plan special
experiments, as    well as the proper theoretical analysis as a
subject of a future work. 

To get a hint why the heating in  the remanent state might be
different from the one in the current-carrying state, we have compared
the power dissipation. According to the estimates, the power
dissipation due to vortex motion is larger in the current-carrying
state. However, in the remanent state there is an extra source for
dissipation -- vortex-antivortex annihilation. Due to this contribution, 
the total power dissipation in the annihilation zone can be larger
than that in the current-carrying state. Consequently the remanent
state can be more unstable with respect to temperature fluctuations that
the current-carrying one.     
It should be noted that a special behavior of vortices in
the vicinity of  the annihilation zone in the remanent state have
already been  addressed  in literature.  The most unusual feature is
meandering instability of flux front   and its turbulent relaxation
observed in very clean single crystals\cite{turbulence}.  The heat
release in the annihilation zone is suggested as a probable
explanation  of this phenomenon. Another explanation  is based upon
the concept of magnetic field concentration  inside the bend of a
current line\cite{vv-holes}.  Furthermore, as discussed in
Ref.~\onlinecite{burlachkov}, near the  annihilation zone, a special
behavior of flux creep can be expected.  It is argued that under
certain assumptions about   $B$-dependence of the pinning energy, the
presence of the annihilation  zone destroys the normal course of flux
creep in the whole sample. 

Another possible source of an instability might be an additional
heat release in the contact pads. However, according to the estimate
given in Sec. IIA, the heat release in contact regions is
negligible. This conclusion is confirmed by the fact that the bridge
burn-out  took place in its central part.\cite{burning} 
 
\section{Conclusion} 
 
Measurements of magnetic flux distribution in a HTSC strip with
transport current, as well as in the remanent state, are performed by
magneto-optical method.  The experimental  results are compared with
predictions of the  critical state model for a strip geometry and with
computer simulation of flux penetration in the flux creep regime.  In
the current-carrying state, the agreement was satisfactory.
Simulation of flux creep predicts slightly deeper flux penetration
than the CSM in agreement with experiment.  In the remanent state
after transport  current, the CSM fails to  explain the experimental
results. Our simulations of flux creep allow  to describe the flux
profiles after relatively small current, $I \le  0.7 I_c$. At larger
currents the total trapped flux appears  substantially less than 
predicted by the flux creep simulations, and it {\em decreases}  with {\em
increasing} $I$.  Excessive power dissipation in the annihilation zone
can be an  explanation of these experimental results. 

Additional experiments and elaborated theoretical models for the
remanent state are under development. We believe that an  important
information can be obtained from time-resolved  studies of remanent
state nucleation.  The conditions for nucleation of macroscopic flux
jumps  is also a subject for future theoretical investigation. 

\acknowledgements
The financial support from the Research Council of Norway and from the
Russian Program for Superconductivity, project No 98031, are grateful
acknowledged.

\widetext 

\begin{references}
\bibitem{end1} The sufficient  condition for Eq.~(\ref{JJtHa}) to be
valid can be written as  $I/I_c \gg B_a/B_c$. The upper estimate for
$B_a$  is  a field generated by two parallel semi-infinite wires
carrying the  current $I$ in the middle of the segment   connecting
their ends, see Fig.~\ref{f_scheme}. Thus $B_a = \eta \mu_0  I/\pi l$,
where  $l$ is the bridge length while  $\eta <1 $ is a  numerical
factor. Our experimental data are compatible with $\eta=0.55$.  Since
$l/w \approx 10$ inequality $I/I_c \gg B_a/B_c$ appears met.
\end{references}
\end{document}